\documentclass[11pt]{article}

\usepackage{cite}
\usepackage{amsmath,amssymb,amsfonts}
\usepackage{algorithm}
\usepackage{algpseudocode}
\usepackage{array}
\usepackage{url}
\usepackage{xcolor}
\usepackage{graphicx}
\usepackage{booktabs}
\usepackage[hidelinks]{hyperref}

\newcommand{\Graph}{\mathcal{G}}
\newcommand{\Vertices}{\mathcal{V}}
\newcommand{\Edges}{\mathcal{E}}
\newcommand{\Advs}{\mathcal{A}}
\newcommand{\Info}{\mathcal{F}}
\newcommand{\Horizons}{\mathcal{D}}
\newcommand{\Indicator}{\mathbb{I}}
\newcommand{\Prob}{\mathbb{P}}

\newcommand{\TimeOnly}{\textsc{TimeOnly}}
\newcommand{\GraphStruct}{\textsc{GraphStruct}}
\newcommand{\HistoryGraph}{\textsc{HistoryGraph}}
\newcommand{\PatchRisk}{\textsc{PatchRisk}}

\begin{document}

\title{PatchRisk: Forecasting Future Vulnerability Exposure
in Open-Source Dependency Networks}

\author{
Ashfaq Ali Shafin\\
\small Department of Mathematics and Computer Science\\
\small Augustana College\\
\small Rock Island, IL, USA\\
\small \texttt{shafinashfaqali21@gmail.com}
\and
Khandaker Mamun Ahmed\\
\small Beacom College of Computer and Cyber Sciences\\
\small Dakota State University\\
\small Madison, SD, USA\\
\small \texttt{khandakermamun.ahmed@dsu.edu}
}

\date{}

\maketitle
\raggedbottom

\begin{abstract}
Open-source software ecosystems are web-scale dependency networks. A downstream package can become exposed to security risk not because its own source code changes, but because one of its transitive dependencies later receives a vulnerability advisory. Existing vulnerability-detection work often focuses on whether code is currently vulnerable or whether a known vulnerable dependency is already present. We study a different problem: future transitive vulnerability exposure. Given a package-version dependency graph observed at release time, the task is to predict whether any non-root dependency will receive a vulnerability advisory within a future horizon.

This problem is important for Web intelligence and software supply-chain security because it supports proactive dependency triage before future exposure is visible. It is also easy to evaluate incorrectly: current vulnerable dependencies can leak the label, and different versions of the same root package can create package-level memorization across train and test splits. We therefore construct \PatchRisk, a leakage-aware benchmark from Open Source Vulnerability (OSV) advisories and deps.dev dependency graphs for npm and PyPI. The benchmark uses filtration-aware labels, package-disjoint evaluation, temporal testing, and nested 1K, 3K, 5K, and 10K sampling scales. The largest cleaned setting contains 9,007 root package-version graphs spanning 4,157 root packages.

We evaluate three feature families: \TimeOnly, \GraphStruct, and \HistoryGraph. On the 10K temporal-test benchmark, \HistoryGraph{} improves AUPRC over the strongest \TimeOnly{} baseline from 0.351 to 0.640 for 90-day forecasting and from 0.473 to 0.813 for 365-day forecasting. The improvement remains stable across smaller scales and package-group shuffle robustness tests. These results show that historical risk in dependency neighborhoods provides meaningful predictive signal for future software supply-chain exposure.
\end{abstract}

\section{Introduction}

Modern software systems are assembled from open-source packages. A web application, data-processing pipeline, mobile backend, or machine-learning service may depend on hundreds or thousands of direct and transitive packages. This creates a large, evolving dependency network in which security risk can propagate across package boundaries. A root package may be safe in isolation at release time, yet later become exposed because one of its transitive dependencies receives a vulnerability advisory. For maintainers and security teams, the practical question is therefore not only whether a package is vulnerable now, but whether its dependency neighborhood indicates elevated future exposure risk.

This paper studies \emph{future transitive vulnerability exposure} as an intelligent risk-forecasting problem over open-source dependency networks. Given a root package version and its dependency graph observed at release time, we predict whether any non-root dependency will receive a vulnerability advisory within a future horizon. We evaluate two horizons, 90 days and 365 days. The 90-day horizon reflects short-term triage: maintainers and security teams may want to know which packages deserve immediate dependency monitoring after release. The 365-day horizon reflects longer-term supply-chain exposure: dependency neighborhoods that repeatedly include historically risky packages may become exposed over a longer time window.

The problem is a natural fit for Web intelligence. Open-source package ecosystems are public, web-accessible, and networked digital infrastructures. They are also socio-technical systems: dependency choices, release practices, maintainer activity, and security advisories interact over time. A forecasting system that combines vulnerability intelligence with dependency-network structure can help prioritize which package-version graphs should receive closer monitoring. Such a system does not replace vulnerability scanners; instead, it complements them by ranking future risk before a future advisory becomes visible.

This task differs from two common vulnerability-analysis settings. First, it differs from code-level vulnerability detection. In code-level detection, the model predicts whether a function, line, file, or repository contains vulnerable code. In our task, the root package may not contain any vulnerable code at all. The positive event occurs when a dependency later receives an advisory. Second, it differs from current-exposure detection. In current-exposure detection, the task is to identify whether a package is already connected to known vulnerable dependencies. In our task, future labels are defined using advisories published after the root package version's release time. Thus, a model must rank packages by future exposure risk rather than simply rediscovering known vulnerabilities.

The problem is deceptively easy to evaluate incorrectly. If current vulnerable dependencies are used as labels, graph features can directly or indirectly reveal the answer. For example, a feature that counts vulnerable dependencies would almost perfectly identify current exposure. Similarly, if different versions of the same root package appear in both train and test sets, models can memorize package-specific dependency signatures rather than generalize to unseen packages. These issues are especially important in dependency ecosystems because popular packages release many versions with similar dependency structure.

We address these issues through leakage-aware benchmark construction. Advisory-derived features are filtered by the root release time. Historical features use only advisories published no later than the root release time. Future labels use only advisories published after the root release time and within the chosen horizon. Evaluation is package-disjoint: all sampled versions of the same root package are assigned to exactly one split. We additionally evaluate a temporal-test setting in which the latest root packages by median release time are held out as test packages.

Our goal is not to introduce a new graph neural network architecture. Instead, we introduce a benchmark and empirical signal characterization for a security-relevant Web intelligence task. We ask whether dependency graph structure and historical risk improve future exposure ranking beyond time-aware baselines. To answer this, we compare three feature families. \TimeOnly{} captures release age and ecosystem identity. \GraphStruct{} adds dependency graph size and structural features. \HistoryGraph{} adds historical advisory counts from non-root dependency nodes. We evaluate these feature families with logistic regression, random forest, and histogram gradient boosting classifiers.

The contributions are:
\begin{itemize}
    \item We formulate future transitive vulnerability exposure as a filtration-aware forecasting task over open-source dependency graphs.
    \item We construct nested public-data benchmarks from OSV and deps.dev over npm and PyPI, scaling from 1K to 10K requested packages. The largest cleaned setting contains 9,007 root package-version graphs and 4,157 root packages.
    \item We introduce package-disjoint and temporal-test evaluation protocols that reduce leakage and measure generalization to unseen packages and future package cohorts.
    \item We show that graph-derived historical-risk features consistently improve AUPRC over time-aware baselines across 90-day and 365-day horizons.
    \item We provide a compact algorithmic description of the \PatchRisk{} benchmark construction and evaluation procedure for reproducibility.
\end{itemize}

\begin{figure}[t]
\centering
\fbox{%
\begin{minipage}{0.94\columnwidth}
\small
\textbf{PatchRisk pipeline.}\\[2pt]
\textbf{1. Vulnerability evidence:} OSV advisories are normalized into ecosystem, package, affected range, and publication time.\\[2pt]
\textbf{2. Dependency structure:} deps.dev is queried for fixed package-version dependency graphs.\\[2pt]
\textbf{3. Leakage-aware labeling:} historical advisories satisfy $\tau\leq t_{p,v}$, while future labels satisfy $t_{p,v}<\tau\leq t_{p,v}+\Delta$.\\[2pt]
\textbf{4. Package-disjoint evaluation:} all versions of the same root package are assigned to one split, and the latest packages are held out for temporal testing.
\end{minipage}}
\caption{Overview of the benchmark construction workflow. The central design principle is that features must be observable at root release time, while labels are future advisory events.}
\label{fig:pipeline}
\end{figure}

\section{Background and Related Work}

\subsection{Open-Source Vulnerability and Dependency Data}

Structured vulnerability data is necessary for package-level forecasting. The Open Source Vulnerability (OSV) schema represents open-source vulnerabilities with package identifiers, affected version ranges, commits, and advisory metadata, enabling vulnerability records to be mapped to package versions~\cite{osv_schema,osv_blog}. OSV is well suited for ecosystem-level analysis because it is designed to represent package-specific affected ranges rather than only free-text vulnerability descriptions. The National Vulnerability Database (NVD) provides standardized CVE metadata and APIs for retrieving vulnerability records at scale~\cite{nvd_api,nvd_feeds}. While NVD is broadly useful, package-ecosystem forecasting benefits from data sources that directly encode affected package versions.

For dependency structure, deps.dev provides package-version metadata and resolved dependency information for open-source ecosystems~\cite{depsdev_api,depsdev_googleblog}. The deps.dev documentation describes dependency graphs as computed artifacts and notes that resolved dependency graphs are ecosystem-dependent approximations~\cite{depsdev_api}. We therefore interpret deps.dev graphs as public fixed-version dependency-resolution artifacts rather than full registry time-travel reconstructions. This distinction matters: our strongest temporal guarantee applies to advisory-derived information. Historical risk features only use advisories published no later than the root release time, while future labels only use advisories published after the root release time.

\subsection{Software Ecosystem Vulnerability Propagation}

Prior empirical software-engineering and security work shows that vulnerabilities propagate through dependency networks and may remain unresolved for long periods. Decan, Mens, and Constantinou studied security reports in the npm dependency network and modeled vulnerable releases, dependency constraints, and propagation to dependent packages~\cite{decan2018impact}. Zimmermann et al. analyzed the npm ecosystem and showed that dependency structure and maintainer concentration create systemic security risks~\cite{zimmermann2019smallworld}. Alfadel et al. studied vulnerabilities affecting Python packages in PyPI~\cite{alfadel2023pypi}. Ponta, Plate, and Sabetta developed methods for detecting, assessing, and mitigating vulnerable open-source dependencies~\cite{ponta2020detection}.

These studies motivate our forecasting task, but our objective is different. Much prior ecosystem work measures known vulnerability impact after advisories are available. We instead ask whether a dependency graph observed for a root package version can predict future exposure before the corresponding dependency advisories are known. This reframing turns software supply-chain exposure into a temporal prediction problem: the model must use release-time graph structure and historical dependency evidence to rank packages by future exposure risk.

\subsection{Graph Learning and Temporal Graph Benchmarks}

Graph representation learning provides methods for prediction over relational data. Random-walk methods such as DeepWalk and node2vec learn representations from graph neighborhoods and enabled scalable prediction on large networks~\cite{perozzi2014deepwalk,grover2016node2vec}. Message-passing graph neural networks such as GCN, GraphSAGE, and GAT aggregate neighborhood information to support node-level or graph-level prediction~\cite{kipf2017gcn,hamilton2017graphsage,velickovic2018gat}. GraphSAGE is particularly relevant because it emphasizes inductive generalization to unseen nodes~\cite{hamilton2017graphsage}. Heterogeneous Graph Transformer extends attention-based graph learning to typed heterogeneous graphs~\cite{hu2020hgt}.

Temporal graph learning is also relevant because package ecosystems evolve over time. DyRep models representation learning over dynamic graphs as a latent process over evolving associations and interactions~\cite{trivedi2019dyrep}. TGAT introduces temporal self-attention and functional time encoding for inductive learning on temporal graphs~\cite{xu2020tgat}. EvolveGCN adapts graph convolutional parameters over graph sequences~\cite{pareja2020evolvegcn}. Causal Anonymous Walks provide inductive representations for temporal networks without relying on node identities~\cite{wang2021caw}. Benchmarking efforts such as the Open Graph Benchmark and Temporal Graph Benchmark emphasize realistic, reproducible evaluation protocols for graph ML across domains~\cite{hu2020ogb,huang2023tgb}. Our work follows this benchmark-driven direction but focuses on software supply-chain risk rather than social, biological, or citation networks.

\subsection{Machine Learning for Software Vulnerability Detection}

A separate line of work applies machine learning to code-level vulnerability detection. VulDeePecker detects vulnerabilities from code gadgets using deep learning~\cite{li2018vuldeepecker}. Devign uses graph neural networks over program semantic representations for function-level vulnerability identification~\cite{zhou2019devign}. LineVul uses transformer-based representations for line-level vulnerability prediction~\cite{fu2022linevul}. These methods focus on whether code contains vulnerable constructs. Our task is different: the root package may become exposed because a dependency later receives an advisory. The prediction unit is therefore a package-version dependency graph rather than a function, line, file, or repository.

\subsection{Positioning}

The closest empirical ecosystem studies quantify vulnerability propagation and dependency risk after advisories are known~\cite{decan2018impact,zimmermann2019smallworld,alfadel2023pypi,ponta2020detection}. The closest graph-learning literature provides general methods and benchmark principles for static, heterogeneous, and temporal graphs~\cite{hamilton2017graphsage,hu2020hgt,xu2020tgat,huang2023tgb}. Our contribution lies at their intersection. We formulate future transitive vulnerability exposure as a filtration-aware temporal graph forecasting problem, construct a package-disjoint benchmark from OSV and deps.dev, and show that historical dependency-risk features improve future exposure ranking over strong time-aware baselines.

\section{Problem Formulation}

Let $(p,v)$ be a root package version released at time $t_{p,v}$. Its dependency graph is a directed graph
\begin{equation}
\Graph_{p,v}=(\Vertices_{p,v},\Edges_{p,v}),
\end{equation}
where the root node is $r=(p,v)$ and an edge $(u,w)\in\Edges_{p,v}$ means package-version node $u$ depends on package-version node $w$. The non-root nodes represent direct and transitive dependencies of the root package version.

Let $\Advs$ denote the set of vulnerability advisories. For dependency node $u\in\Vertices_{p,v}$, let
\begin{equation}
\mathcal{T}_{u}=\{\tau_a: a\in\Advs,\; a \text{ affects package-version } u\},
\end{equation}
where $\tau_a$ is the advisory publication time. For a forecasting horizon $\Delta$, define historical and future advisory sets relative to the root release time:
\begin{equation}
\mathcal{T}^{-}_{u}(t_{p,v}) = \{\tau\in\mathcal{T}_{u}:\tau\leq t_{p,v}\},
\end{equation}
\begin{equation}
\mathcal{T}^{+}_{u}(t_{p,v},\Delta) = \{\tau\in\mathcal{T}_{u}:t_{p,v}<\tau\leq t_{p,v}+\Delta\}.
\end{equation}

The future transitive exposure label is
\begin{equation}
y_{p,v}^{(\Delta)} =
\Indicator\left[\exists u\in\Vertices_{p,v}\setminus\{r\}: \mathcal{T}^{+}_{u}(t_{p,v},\Delta)\neq\emptyset\right].
\end{equation}
This label is positive if at least one non-root dependency receives an advisory after the root release time and within the forecast horizon. Advisories published at or before the root release time are excluded from the future label.

We define two historical dependency-risk features:
\begin{equation}
H_{count}(p,v)=\sum_{u\in\Vertices_{p,v}\setminus\{r\}}\Indicator[\mathcal{T}^{-}_{u}(t_{p,v})\neq\emptyset],
\end{equation}
\begin{equation}
H_{events}(p,v)=\sum_{u\in\Vertices_{p,v}\setminus\{r\}}|\mathcal{T}^{-}_{u}(t_{p,v})|.
\end{equation}
$H_{count}$ counts how many dependency nodes had at least one historical advisory by the root release time. $H_{events}$ counts the total number of historical dependency advisories. These features are intentionally simple and auditable: they represent observable historical risk in the dependency neighborhood.

The forecasting task is to estimate
\begin{equation}
f_\theta(\Graph_{p,v},\Info_{t_{p,v}})
\approx
\Prob(y_{p,v}^{(\Delta)}=1\mid \Graph_{p,v},\Info_{t_{p,v}}),
\end{equation}
where $\Info_{t_{p,v}}$ is the information available by the root release time. The filtration constraint is central: a valid feature may depend on dependency structure and historical advisories, but not on advisories that occur after the root release time.

\section{PatchRisk Benchmark Construction}

\subsection{Design Goals}

The \PatchRisk{} benchmark is designed around four goals.

\textbf{Temporal validity.} Labels must represent future events relative to the root release time. A model should not be rewarded for detecting already-known vulnerable dependencies. This requirement is especially important because open-source advisories often describe vulnerabilities in old package versions, while the public advisory appears much later.

\textbf{Package-level generalization.} Evaluation should measure generalization to unseen root packages, not only unseen versions. Therefore, all sampled versions of the same root package are assigned to the same split. This avoids a common shortcut in package-version prediction: learning stable package identity or package-specific dependency signatures.

\textbf{Public-data reproducibility.} The benchmark uses publicly available OSV advisory data and deps.dev dependency graph data. This makes the task reproducible and allows future work to improve the pipeline with additional ecosystems or more precise dependency-resolution snapshots.

\textbf{Multi-scale robustness.} Results should not depend on a single sample size. We therefore evaluate nested sampling scales: 1K, 3K, 5K, and 10K requested packages. The smaller scales act as robustness checks, while the 10K setting is the primary benchmark because it contains the largest package diversity.

\subsection{Algorithm}

Algorithm~\ref{alg:patchrisk} summarizes the full \PatchRisk{} benchmark pipeline. Line~1 first normalizes OSV advisories into affected package-version records, so that vulnerabilities can be matched consistently by ecosystem, package, affected version range, and advisory time. Lines~2--4 define the multi-scale sampling procedure: for each scale, the algorithm selects root packages from npm and PyPI and keeps at most $K$ temporally spaced versions per package to avoid overrepresenting packages with many releases.

Lines~5--8 construct the dependency-graph representation for each sampled root version. For each root package version $(p,v)$, the algorithm queries deps.dev, obtains the dependency graph $G_{p,v}$, records the root release time $t_{p,v}$, and computes structural graph features. Lines~9--18 then construct leakage-aware labels and historical-risk features. Historical features use only advisories with $\tau \leq t_{p,v}$, while future labels are set only when a dependency receives an advisory in the interval $t_{p,v}<\tau\leq t_{p,v}+\Delta$. This separation ensures that the model does not use future advisory information as input.

Line~19 stores the three feature views used in the experiments: \TimeOnly, \GraphStruct, and \HistoryGraph. Lines~21--22 enforce package-disjoint temporal evaluation by splitting data so that no root package appears in multiple splits and by holding out the latest packages as the temporal test set. Finally, lines~23--29 train models for each forecast horizon and feature view, tune the classification threshold on validation data, and report AUROC, AUPRC, F1, and balanced accuracy on the held-out test set.
\begin{algorithm}[t]
\caption{\PatchRisk{} Benchmark Construction and Evaluation}
\label{alg:patchrisk}
\footnotesize
\begin{algorithmic}[1]
\Require OSV advisories $\Advs$; deps.dev package-version records $P$; horizons $\Horizons={90,365}$; scales $S$; max versions per package $K$
\Ensure Test metrics for \TimeOnly, \GraphStruct, and \HistoryGraph{}

\State Normalize $\Advs$ into affected package-version records $(e,p,r,\tau)$

\ForAll{scale $s \in S$}
\State $R_s \gets$ sample up to $s$ npm/PyPI root packages
\State $V_s \gets$ select at most $K$ temporally spaced versions per root package
\ForAll{root version $(p,v) \in V_s$}
    \State $G_{p,v} \gets$ deps.dev dependency graph for $(p,v)$
    \State $t_{p,v} \gets$ release time of root version $(p,v)$
    \State Compute structural features from $G_{p,v}$
    \State Initialize historical-risk features and labels $y^{90}_{p,v}=0$, $y^{365}_{p,v}=0$

    \ForAll{dependency $u \in G_{p,v}$}
        \State Match $u$ to affected OSV records
        \State Add advisories with $\tau \leq t_{p,v}$ to historical-risk features

        \ForAll{$\Delta \in \Horizons$}
            \If{some matched advisory has $t_{p,v} < \tau \leq t_{p,v}+\Delta$}
                \State Set $y^{\Delta}_{p,v}=1$
            \EndIf
        \EndFor
    \EndFor

    \State Store feature views $X_{\textsc{Time}}$, $X_{\textsc{Graph}}$, $X_{\textsc{HistGraph}}$ and $y^{90},y^{365}$
\EndFor

\State Create package-disjoint train, validation, and test splits
\State Hold out latest packages as the temporal test set

\ForAll{$\Delta \in \Horizons$}
    \ForAll{$X \in \{X_{\textsc{Time}},X_{\textsc{Graph}},X_{\textsc{HistGraph}}\}$}
        \State Train classifier using feature view $X$
        \State Tune threshold on validation examples
        \State Evaluate AUROC, AUPRC, F1, and balanced accuracy on test examples
    \EndFor
\EndFor
\EndFor
\end{algorithmic}
\end{algorithm}

\subsection{Feature Views}

We evaluate three feature views.

\textbf{\TimeOnly.} This view contains root release age and ecosystem identity. It captures the possibility that older packages, newer packages, or particular ecosystems have different baseline exposure rates. Because temporal effects can be strong in vulnerability data, \TimeOnly{} is a necessary baseline. It also prevents us from overstating the contribution of graph features when the gain could be explained by simple time or ecosystem effects.

\textbf{\GraphStruct.} This view augments \TimeOnly{} with dependency graph structure: number of nodes, number of edges, direct dependency count, transitive dependency count, and graph density. These features test whether graph size and connectivity alone are predictive of future exposure. A package with more dependencies has more possible exposure points, but graph size alone does not reveal whether those dependencies have historically risky advisory patterns.

\textbf{\HistoryGraph.} This view augments \GraphStruct{} with historical dependency-risk features: $H_{count}$ and $H_{events}$. These features test whether prior vulnerability history in the dependency neighborhood helps forecast future exposure. The features are deliberately simple because the primary benchmark question is whether leakage-aware historical graph signals matter, not whether a complex model can exploit obscure correlations.

The feature design intentionally avoids post-release advisory information. In particular, we exclude features that directly count future vulnerable dependencies, shortest paths to future vulnerable nodes, or other quantities that would reveal the label.

\section{Experimental Setup}

\subsection{Data Sources and Normalization}

We parse OSV records for npm and PyPI and normalize affected package entries into advisory ID, ecosystem, package, publication time, and affected-version fields. Publication time is used as the observation time of an advisory. This choice is operationally meaningful: security tools and maintainers can only react to advisories once they become publicly observable. However, advisory publication time is not necessarily the same as vulnerability introduction time. We return to this point in the threats-to-validity section.

We then query deps.dev for package-version metadata and resolved dependency graphs. For each selected root package version, we retain the root release time and the dependency graph returned by deps.dev. We construct root package-version examples only when the required release-time and dependency-graph metadata can be recovered.

\subsection{Sampling Strategy}

Package ecosystems are highly skewed. A small number of popular packages may have many versions and large dependency neighborhoods. If package versions are sampled naively, the benchmark may become dominated by repeated versions of a small number of packages. To reduce this concentration, we select root packages and sample at most $K=5$ temporally spaced versions per root package. This creates a more diverse set of root packages while preserving multiple release-time observations for some packages.

We evaluate four nested sampling scales: 1K, 3K, 5K, and 10K requested root packages. After metadata recovery, dependency graph construction, and label construction, the cleaned datasets have different numbers of usable root package-version graphs. Table~\ref{tab:scale} summarizes the temporal benchmark scale.

\begin{table}[t]
\centering
\caption{Temporal benchmark scale. All splits are package-disjoint. The 10K setting uses a future temporal test set with stratified package-disjoint validation from earlier packages.}
\label{tab:scale}
\scriptsize
\begin{tabular}{lrrrrrr}
\hline
Scale & Examples & Packages & Train & Val. & Test & Test pkg. \\
\hline
1K & 1,113 & 321 & 800 & 106 & 207 & 64 \\
3K & 5,572 & 1,588 & 4,067 & 436 & 1,069 & 317 \\
5K & 6,405 & 2,277 & 4,675 & 257 & 1,473 & 455 \\
10K & 9,007 & 4,157 & 6,040 & 821 & 2,146 & 831 \\
\hline
\end{tabular}
\end{table}

\subsection{Labels and Forecast Horizons}

For each root package-version graph, we consider all non-root dependency nodes. A root graph receives a positive label for horizon $\Delta$ if at least one dependency node matches an OSV affected package-version record whose advisory publication time falls in $(t_{p,v},t_{p,v}+\Delta]$. We evaluate $\Delta=90$ days and $\Delta=365$ days.

The two horizons serve different purposes. The 90-day horizon is a short-term warning signal: it asks whether a dependency neighborhood will soon become exposed. The 365-day horizon is a longer-term supply-chain risk signal: it asks whether the dependency neighborhood is structurally or historically associated with exposure over the following year.

\subsection{Models and Hyperparameters}

For each feature family, we train three standard classifiers: logistic regression, random forest, and histogram gradient boosting. We also include a majority baseline. The purpose is not to exhaustively tune algorithms, but to compare feature families under a controlled evaluation protocol. Logistic regression provides a simple linear baseline. Random forest and histogram gradient boosting provide nonlinear baselines that can capture interactions between graph size, ecosystem, release age, and historical advisory counts.

Numeric features are median-imputed and standardized where appropriate. Ecosystem is treated as a categorical feature. Logistic regression uses class balancing because the positive class is rare. Tree-based models are trained with fixed random seeds. Decision thresholds are selected on validation data by maximizing F1 and then applied without modification to the test split.

\subsection{Metrics}

We report AUROC, area under the precision-recall curve (AUPRC), F1, and balanced accuracy. Because future exposure is imbalanced, AUPRC is the primary metric. AUPRC directly measures ranking quality under class imbalance and is more informative than accuracy when positive examples are sparse. F1 and balanced accuracy are reported as thresholded metrics. The decision threshold is selected on validation data and then applied to test data.

\subsection{Evaluation Protocols}

All evaluations are package-disjoint. This means no root package appears in more than one split, even if multiple versions of that package are sampled. Package-disjoint splitting is stricter than version-level splitting and reduces package-level memorization.

We use two protocols. \textbf{Package-group shuffle} assigns root packages randomly to train, validation, and test sets. This protocol measures generalization to unseen packages under a random package split. \textbf{Temporal-test evaluation} orders root packages by median observed root-version release time and holds out the latest packages as the temporal test set. This protocol measures generalization to future package cohorts. In the 10K setting, the middle temporal validation segment contained zero positive examples. We therefore preserved the future temporal test set and selected a package-disjoint stratified validation set from earlier packages. This keeps the test set temporal while allowing threshold tuning on a validation split with both classes.

\section{Results}

\subsection{RQ1: Does Historical Dependency Risk Improve Temporal Forecasting?}

Table~\ref{tab:temporal_ap} reports temporal evaluation across scales and horizons. \HistoryGraph{} consistently improves AUPRC over \TimeOnly. The 10K benchmark is the main result because it has the largest package diversity. In the 10K temporal-test setting, \HistoryGraph{} improves AUPRC from 0.351 to 0.640 for 90-day forecasting and from 0.473 to 0.813 for 365-day forecasting.

\begin{table}[t]
\centering
\caption{Temporal evaluation across scales. Values are test-set AUPRC except relative gain, which compares \HistoryGraph{} to \TimeOnly.}
\label{tab:temporal_ap}
\scriptsize
\begin{tabular}{llrrrrr}
\hline
Scale & Horizon & Maj. & Time & Struct. & Hist. & Gain \\
\hline
1K & 90d & 0.159 & 0.583 & 0.536 & 0.629 & 8.0\% \\
1K & 365d & 0.266 & 0.655 & 0.777 & 0.804 & 22.7\% \\
3K & 90d & 0.107 & 0.411 & 0.592 & 0.648 & 57.9\% \\
3K & 365d & 0.202 & 0.507 & 0.769 & 0.777 & 53.2\% \\
5K & 90d & 0.107 & 0.404 & 0.593 & 0.636 & 57.3\% \\
5K & 365d & 0.208 & 0.517 & 0.799 & 0.820 & 58.7\% \\
10K & 90d & 0.104 & 0.351 & 0.551 & 0.640 & 82.5\% \\
10K & 365d & 0.212 & 0.473 & 0.795 & 0.813 & 72.0\% \\
\hline
\end{tabular}
\end{table}

The results show that time and ecosystem identity are informative but incomplete. \TimeOnly{} already performs substantially above the majority baseline, indicating that exposure risk is not temporally uniform. However, graph-derived historical risk provides a large additional gain. The gain is especially important because the task is a ranking problem: maintainers or security teams may use the model to prioritize which package dependency neighborhoods deserve closer monitoring.

\subsection{RQ2: How Strong is the Largest 10K Benchmark?}

Table~\ref{tab:tenk_detail} gives detailed 10K temporal-test results. For 90-day forecasting, \HistoryGraph{} achieves the best score on every metric. For 365-day forecasting, \HistoryGraph{} achieves the best AUROC and AUPRC, while \GraphStruct{} has slightly higher F1 and balanced accuracy. Because future exposure is imbalanced and the operational use case is triage, AUPRC remains the primary metric.

\begin{table}[t]
\centering
\caption{Detailed 10K temporal-test results. AUPRC is the primary metric because future exposure is imbalanced.}
\label{tab:tenk_detail}
\scriptsize
\begin{tabular}{llrrrr}
\hline
Horizon & Family & AUROC & AUPRC & F1 & Bal. Acc. \\
\hline
90d & Majority & 0.500 & 0.104 & 0.188 & 0.500 \\
90d & \TimeOnly & 0.850 & 0.351 & 0.377 & 0.751 \\
90d & \GraphStruct & 0.908 & 0.551 & 0.530 & 0.753 \\
90d & \HistoryGraph & 0.924 & 0.640 & 0.594 & 0.763 \\
\hline
365d & Majority & 0.500 & 0.212 & 0.350 & 0.500 \\
365d & \TimeOnly & 0.852 & 0.473 & 0.611 & 0.825 \\
365d & \GraphStruct & 0.938 & 0.795 & 0.713 & 0.852 \\
365d & \HistoryGraph & 0.944 & 0.813 & 0.710 & 0.801 \\
\hline
\end{tabular}
\end{table}

These results indicate that dependency graph information is useful at both short and long horizons. The 90-day task is more difficult because positives are rarer and the forecast window is shorter. Even in this harder setting, \HistoryGraph{} improves AUPRC by 82.5\% relative to the strongest \TimeOnly{} baseline. The 365-day task has a higher positive rate and stronger overall performance, but the relative gain from \HistoryGraph{} remains large.

\subsection{RQ3: Are the Results Robust Across Scale?}

The multi-scale design helps determine whether the signal is a small-sample artifact. Table~\ref{tab:temporal_ap} shows that \HistoryGraph{} is consistently strong from 1K through 10K. The exact AUPRC values vary by scale because the class distribution and package composition change, but the qualitative ordering is stable: \TimeOnly{} improves over majority, \GraphStruct{} usually improves over \TimeOnly, and \HistoryGraph{} is the strongest feature family by AUPRC in the main temporal setting.

The 1K setting is useful as a small pilot, but it has limited package diversity. The 3K and 5K settings provide intermediate robustness checks. The 10K setting is the most credible main benchmark because it contains the largest number of unique root packages and test packages. We therefore report 10K as the main result and use the smaller scales to show that the trend is not unique to a single sample.

\subsection{RQ4: Does the Pattern Hold Under Package-Group Shuffle?}

Table~\ref{tab:shuffle_ap} reports package-group shuffle robustness. This protocol is not a replacement for temporal testing, but it provides a complementary check under random package-disjoint splits. The same qualitative pattern appears: \HistoryGraph{} outperforms \TimeOnly{} across all scales and horizons. This suggests that the predictive signal is not caused solely by the temporal split design.

\begin{table}[t]
\centering
\caption{Package-group shuffle robustness. Values are test-set AUPRC except relative gain, which compares \HistoryGraph{} to \TimeOnly.}
\label{tab:shuffle_ap}
\scriptsize
\begin{tabular}{llrrrrr}
\hline
Scale & Horizon & Maj. & Time & Struct. & Hist. & Gain \\
\hline
1K & 90d & 0.084 & 0.503 & 0.608 & 0.705 & 40.0\% \\
1K & 365d & 0.155 & 0.684 & 0.728 & 0.853 & 24.8\% \\
3K & 90d & 0.070 & 0.427 & 0.583 & 0.639 & 49.8\% \\
3K & 365d & 0.128 & 0.496 & 0.742 & 0.769 & 55.0\% \\
5K & 90d & 0.053 & 0.335 & 0.667 & 0.771 & 129.8\% \\
5K & 365d & 0.098 & 0.390 & 0.697 & 0.771 & 97.6\% \\
10K & 90d & 0.045 & 0.324 & 0.540 & 0.638 & 96.7\% \\
10K & 365d & 0.092 & 0.422 & 0.751 & 0.761 & 80.2\% \\
\hline
\end{tabular}
\end{table}

The shuffle setting also highlights the importance of class imbalance. The majority baseline AUPRC is low, especially for 90-day prediction, because future exposure is rare. In such settings, small improvements in ranking quality can be operationally meaningful. The observed improvements are not small: at 10K scale, \HistoryGraph{} nearly doubles the 90-day AUPRC relative to \TimeOnly{} under package-group shuffle.

\section{Discussion}

\subsection{Graph History Matters Beyond Time}

The results show that temporal information is important but insufficient. \TimeOnly{} is a strong baseline because vulnerability reporting, package age, and ecosystem-specific practices vary over time. However, dependency graph history adds substantial signal. This suggests that risk is not distributed uniformly across dependency neighborhoods. Packages that depend on historically risky dependency neighborhoods may be more likely to become exposed again in the future.

This observation is consistent with the intuition that open-source ecosystems contain persistent risk concentration. Some dependency neighborhoods may include packages with large attack surfaces, high maintenance burden, broad transitive reach, or repeated vulnerability histories. \PatchRisk{} does not claim to identify the causal mechanism behind every future advisory. Instead, it shows that simple historical graph features are predictive under leakage-aware evaluation.

\subsection{Why Package-Disjoint Evaluation Matters}

During benchmark development, less restrictive evaluation produced overly optimistic results. This is expected in package-version data. If versions of the same package appear in both train and test sets, models can learn stable package-specific dependency signatures. Such results may overstate generalization to unseen packages. Package-disjoint evaluation is therefore necessary for a credible benchmark.

Temporal testing is also important. Random package splits are useful robustness checks, but they do not fully measure deployment-like forecasting. A security team wants to rank future or newly observed packages using models trained on earlier evidence. Holding out the latest packages as temporal test data better reflects this operational setting.

\subsection{Why AUPRC is the Primary Metric}

Future exposure is imbalanced. In the 10K temporal-test setting, the 90-day test positive rate is about 10.4\%, while the 365-day test positive rate is about 21.2\%. Accuracy would be misleading because a model could perform well by predicting the majority class. AUROC is useful but can also look optimistic under class imbalance. AUPRC focuses on ranking positives ahead of negatives and is therefore more aligned with security triage. If a maintainer can only inspect a limited number of packages, the ranking quality of the model is more important than raw accuracy.

\subsection{Practical Implications}

\PatchRisk{} can support proactive dependency monitoring. A package manager, security platform, or maintainer dashboard could use future exposure scores to prioritize dependency audits. For example, a package-version graph with many historically risky transitive dependencies may deserve earlier review, stricter update monitoring, or additional dependency pinning analysis. The benchmark does not replace vulnerability scanners. Instead, it complements them by forecasting exposure before the future advisory is visible.

The results also suggest that dependency-risk tools should consider graph context. Simple time-aware baselines are informative, but dependency-neighborhood history provides stronger ranking performance. This supports the broader Web intelligence view that open-source software ecosystems should be modeled as evolving networks rather than isolated packages.

\section{Threats to Validity}

\subsection{Dependency Graph Time Consistency}

We retrieve dependency graphs for fixed root package versions from deps.dev. This is a practical approximation to release-time dependency structure, but it is not a complete registry time-travel reconstruction. Dependency resolution can depend on ecosystem rules, version constraints, lockfiles, platform assumptions, and later package availability. We therefore avoid claiming that deps.dev gives perfect historical installation snapshots. Our strongest temporal guarantee is instead applied to advisory-derived features: historical risk uses advisories published no later than the root release time, and future labels use advisories published after the root release time.

\subsection{Advisory Timestamp Semantics}

We use advisory publication time as the observation time. This is appropriate for an operational forecasting task because public advisories are what security teams can observe. However, a vulnerability may exist before it is publicly reported, and OSV records may be edited after publication. Therefore, the target is future known advisory exposure, not the unobservable time at which a latent vulnerability first existed. This distinction should be considered when interpreting the results.

\subsection{OSV Matching and Version Ranges}

Mapping dependency nodes to OSV affected package-version records requires ecosystem, package, and version matching. Version-range semantics differ across ecosystems, and dependency graph nodes may not always provide complete information. Imperfect matching can introduce false negatives or false positives in labels and historical-risk features. This is a common challenge in ecosystem-scale vulnerability studies and motivates future work on more precise version-range resolution.

\subsection{Sampling Bias}

The benchmark focuses on npm and PyPI and samples a bounded number of versions per package. This improves diversity but may underrepresent packages with many releases or ecosystems with different dependency-resolution behavior. The nested 1K, 3K, 5K, and 10K scales reduce the risk that results depend on one sample size, but they do not eliminate all sampling bias. Future work should extend the benchmark to additional ecosystems such as Maven, crates.io, Go, and NuGet.

\subsection{Model Scope}

We intentionally use strong non-neural baselines rather than proposing a new GNN architecture. This makes the benchmark easier to interpret and reduces the risk of hiding leakage behind complex models. However, more expressive temporal graph models may improve performance if applied carefully. Future work can use \PatchRisk{} as a benchmark for graph neural networks, temporal graph models, survival models, and calibrated risk-ranking systems.

\section{Ethics and Responsible Use}

This work uses public vulnerability and package metadata for defensive prioritization. We do not release exploit code, exploitation instructions, or private vulnerability information. The intended use is to help maintainers, security teams, and ecosystem operators prioritize dependency monitoring and proactive audits.

Risk forecasting can also be misused if interpreted as a definitive judgment about a package or maintainer. A high predicted exposure score does not mean that a package is malicious or currently vulnerable. It means that the dependency neighborhood resembles patterns associated with future advisory exposure. Such scores should be used as triage signals, not as automatic blocking decisions without human review.

Responsible deployment should include transparency, calibration, and appeal mechanisms. Maintainers should be able to understand why a package was flagged, and security teams should combine predictive signals with conventional vulnerability scanning, dependency update policies, and human expertise.

\section{Conclusion}

We introduced future transitive vulnerability exposure forecasting as a Web-intelligence and graph-mining problem over open-source dependency networks. Using OSV advisories and deps.dev dependency graphs, we built nested benchmarks from 1K to 10K requested packages. The largest cleaned benchmark contains 9,007 root package-version graphs across 4,157 root packages. Under package-disjoint temporal-test evaluation, graph-derived historical-risk features substantially improve AUPRC over time-only baselines for both 90-day and 365-day horizons. The results are robust across smaller scales and package-group shuffle evaluation. These findings show that dependency-neighborhood history carries predictive signal for future software supply-chain exposure and provide a leakage-aware benchmark for intelligent vulnerability-risk forecasting.

\bibliographystyle{IEEEtran}
\bibliography{references}

\end{document}